\documentstyle[12pt]{article}
\def\be{\begin{equation}}       \def\ba{\begin{array}}
\def\ee{\end{equation}}         \def\ea{\end{array}}
\def\bea {\begin{eqnarray}}      \def\eea {\end{eqnarray}}
\def\bean{\begin{eqnarray*}}    \def\eean{\end{eqnarray*}}

\def\const {\mathop{\rm const}\nolimits}

\def\<{\langle} \def\({\left(}  \def\>{\rangle} \def\){\right)}
\def\defeq {\stackrel{\mbox{\rm\small def}}{=}}
\newtheorem{exi}{Example}

\title{A new class of linearizable equations}
\author{R. Hern\'andez Heredero, A. Shabat and V. Sokolov}

\begin{document}
\maketitle
\thispagestyle{empty}

\begin{abstract} Using the symmetry approach, we find a class of integrable
nonlinear PDEs with dispersion law $\omega(k)=k^{\frac32}$. All these
equations turn out to be linearizable by means of a differential
parametrization.
\end{abstract}

\section{Introduction}
We consider the  problem of classification of PDEs
\be \label{qF} q_{tt}=q_{xxx}+F(q,q_x,q_t,q_{xx},q_{tx}).             \ee
having higher symmetries. These equations
possess an unusual dispersion law $\omega(k)=k^{\frac32}.$ Moreover one can
prove that for such equations there are no higher conserved densities
and therefore they are nonhamiltonian.

The first example
\be\label{311}
q_{tt}=q_{xxx}+3q_xq_{tx}+(q_t-3q_x^2)q_{xx}
\ee
of such equations has been found in \cite{luis3}.
Particular solutions of (\ref{311}) feature moving branch-points
singularities.
For instance, the simplest self-similar solutions of the form $q=q(x+kt)$
satisfy the ODE
\[y'=y(y-k)^2+\const \]
where $q_x=y$. It is obvious that they have singularities of such type.

Nevertheless, equation (\ref{311}) possesses a degenerate
Lax representation which implies infinitely many higher symmetries
and corresponding invariant solutions \cite{luis3}. Three higher symmetries
are as follows
\bean
 && q_{t_1}=q_{xx}+2q_x q_t-q_x^3\\
 && q_{t_2}=q_{xt}+q_x q_{xx}+q_t^2+q_x^2 q_t-q_x^4\\
 && q_{t_3}=q_{tt}+3q_x q_t^2-2 q_x^3 q_t.
\eean
The whole hierarchy of symmetries can be reproduced by a
recursion operator
\begin{equation}\label{r0}
{\cal R}_{0}=-q_{x}+2 D^{-1} q_{xx}+D^{-1} D_{t}=A+B D_{t}
\end{equation}
acting on the seed symmetries $q_x$ and $q_t$.

In Section 2 we generalize the main concepts of the symmetry approach, such
as
the formal recursion operator and canonical conserved densities for the
case of non-evolutionary equations of the form
\be\label{qtt}
q_{tt}=F(q,q_1,q_2,\dots , q_n,\, q_{t},q_{t1}, q_{t2},\dots,q_{tm}).
\ee
Such type of equations were excluded from consideration in works
\cite{is,ss,dss},
where only evolution equations were investigated. Obviously, equation
(\ref{qtt}) can be rewritten as a system of two evolution equations. For
example, equation (\ref{311}) is equivalent to
\[u_{t}=(v_{x}-u \,v)_{x},\qquad
v_{t}=(u-v^{2})_{x},    \]
where $v=q_{x}$. However the matrix coefficients of leading derivatives for
such systems
have the structure of a Jordan block, whereas in papers
\cite{msy,svin1,svin2},
devoted to systems of evolution equations, the leading matrix was supposed
to be diagonalizable.

Using the technique of canonical conserved densities, we find all equations
(\ref{qF}) possessing infinitely many higher symmetries.
It turns out (see Section 3) that all these equations are related to second
order evolution equations
\begin{equation} \label{svin}
u_{t}=H(x,u,u_{x},u_{xx})
\end{equation}
having higher symmetries. It was shown by Sergey Svinolupov \cite{Sv} that
any such equation~(\ref{svin}) can by reduced to one of the following
equations
$$
u_{t}=u_{xx}+f(x) u,
$$
$$
u_{t}=u_{xx}+2 u u_{x}+g(x),
$$
$$
u_{t}=\Big(\frac{u_{x}}{u^{2}}+\lambda x\Big)_{x},
$$
$$
u_{t}=\Big(\frac{u_{x}}{u^{2}}+\lambda_{1} x u+\lambda_{2} u\Big)_{x}
$$
by a contact (or point) transformation
\begin{equation}\label{cont}
\bar x=\phi(x,u,u_{x}), \qquad \bar u=\psi(x,u,u_{x}), \quad
\end{equation}
where
$$
\frac{\partial \phi}{\partial u_{x}} \left(\frac{\partial \psi}{\partial u}
u_{x}+
\frac{\partial \psi}{\partial x}\right)=\frac{\partial \psi}{\partial u_{x}}
\left(\frac{\partial \phi}{\partial u} u_{x}+
\frac{\partial \phi}{\partial x}\right).
$$
All these equations can be linearized by simple differential substitutions
(see \cite{Sv}).

In Section 3 we show that all equations (\ref{qF}) from our list admit a
parametrization of the form
\begin{equation} \label{linsys}
q_{x}=K(q,q_{y},q_{yy}), \qquad q_{t}=S(q,q_{y},q_{yy},q_{yyy}),
\end{equation}
where $q_{x}=K$ is a linearizable equation of second order and $q_{t}=S$ is
a higher symmetry of this equation.  For example, the parametrization
(\ref{linsys}) of equation (\ref{311}) is given by \[
q_x=\frac{q_{yy}}{q_y^2},\qquad
q_t=-\frac{q_{yyy}}{q_y^3}+3\frac{q^2_{yy}}{q_y^4}.
\]

Any common solution $q(x,t,y)$ of system
(\ref{linsys}) gives us an one-parameter family of solutions for the integrable
equation (\ref{qF}). However, the general solution of (\ref{linsys}) depends
on
one function $q_{0}(y)=q(0,0,y)$ of one variable, whereas the general
solution of (\ref{qF}) depends on two functions of one variable.

Although a general idea of such parametrization is contained in
\cite{luis1,luis3}, explicit formulas for
linearization of equation (\ref{311}) were obtained first by V. E. Adler.

{\bf Acknowledgements.} The authors are grateful to V. Adler
for useful discussions. This research was partly supported by the Russian
Fund for Basic Research (grants 01-01-00874-A, 02-01-00431-A and
00-15-96007-L) and Spanish project DGICYT PB98-0821.

\section{Classification of third order equations.}

\subsection{Integrability conditions.}

For equation (\ref{qtt}) all mixed derivatives of $q$ containing at
least two time differentiation can be expressed in terms of
\begin{equation} \label{dynvar}
q, \, q_{x},\, q_{xx},\, \dots, q_{i}, \dots,  \qquad q_{t},\,
q_{t1}=q_{tx},\,
q_{t2}=q_{txx},\,\dots, q_{ti}, \dots
\end{equation}
in virtue of (\ref{qtt}).
The derivatives (\ref{dynvar}) are regarded as {\it independent} variables.

An equation
\be\label{rs}
q_\tau=G(q,q_1,q_2,\dots , q_{r},\,\, q_{t},q_{t1}, q_{t2},\dots,q_{ts})
\ee
compatible with (\ref{qtt}) is called infinitesimal (local) symmetry of
(\ref{qtt}). Compatibility implies that the function $G$ satisfies the
equation ${\cal F}(G)=0$, where
$$
{\cal F}=D_{t}^{2}-\sum_{i=0}^{n} \frac{\partial F}{\partial
q_{i}}\,D_{x}^{i}-
\left(\sum_{i=0}^{m} \frac{\partial F}{\partial q_{ti}}\,D_{x}^{i}\right)
D_{t}\defeq D_{t}^{2}-(M+ND_t)
$$
is the linearization operator for equation (\ref{qtt}).

In order to rewrite consistency conditions of (\ref{qtt}) and (\ref{rs})
in terms of a series of conservation laws
\begin{equation}\label{cond}
(\rho_{i})_{t}=(\sigma_{i})_{x}
\end{equation}
for (\ref{qtt}), one can use a formal
Lax representation of the problem. The linearization of equations
(\ref{qtt}),
(\ref{rs}) gives rise to the compatibility problem for linear equations
\[ \phi_{tt}=(M+ND_t)\,\phi,\qquad  \phi_\tau=(A+BD_t)\,\phi     \]
or, equivalently,
\[ \Phi_t=F_*\Phi,\quad  \Phi_\tau=G_*\Phi,\qquad \Phi=\left(\ba{c}\phi\\
                            \phi_t\ea\right),
\quad F_*=\left(\ba{cc}   0 & 1\\
                      M & N\ea\right)
\]
where
\be\label{G*}  G_*=
   \left(\ba{cc}  A & B\\
           \hat A & \hat B\ea\right),\quad \hat A\defeq A_t+BM ,
           \quad \hat B\defeq B_t+BN+A
\ee
The cross differentiation yields
 \be\label{GF} D_t(G_*)=[F_*,G_*]+D_\tau(F_*)   \ee
where $F_*,\, G_*$ are matrix differential operators. The crucial
step in the symmetry approach (see \cite{ss, msy, mss} and references there)
is to consider instead of above equation one as follows
\be\label{RF}
  D_t(R)=[F_*,\,R],
\ee
where $R$ is matrix pseudo-differential operator. We call $R$
{\it matrix formal recursion operator.}

Denoting as before $R_{11}=A,\, R_{12}=B$ we can rewrite (\ref{RF}) as
follows
\be\label{*}
A_{tt}-NA_t+[A,M]+(2B_t+[B,N])M+BM_t=0
\ee
\be\label{**}
B_{tt}+2A_t+[B,M]+[A,N]+([B,N]+2B_t)N+BN_t-NB_t=0.
\ee
If $R_1,\, R_2$ are formal matrix recursion operators, then $R_3 = R_1R_2$
is a formal recursion operator as well and we find using (\ref{G*}) that
 \be\label{***}
 A_3=A_1A_2+B_1B_2M+B_1A_{2,t},\qquad B_3=A_1B_2+B_1A_2+B_1B_2N+B_1B_{2,t}.
 \ee

Identities (\ref{*}),(\ref{**}) mean that the scalar pseudo-differential
operator
${\cal R}=A+B\,D_{t}$ is related to the linearization ${\cal F}$
of equation (\ref{qtt}) by
\begin{equation} \label{defrec}
{\cal F} (A+B\,D_{t})=(\bar A+B\,D_{t}) {\cal F},
\end{equation}
where $\bar A=A+2 B_{t}+[B,N]$.
The pseudo-differential operator ${\cal R}=A+B\,D_{t}$, whose
components
 $$
 A= \sum_{-\infty}^{n} a_{i} D_{x}^{i}, \qquad
 B= \sum_{-\infty}^{m} b_{i} D_{x}^{i}
 $$
satisfy (\ref{*}) and (\ref{**}), is called
{\it scalar formal recursion operator} for
equation (\ref{qtt}). If $A$ and $B$ are differential operators
(or ratios of differential operators), condition (\ref{defrec})
implies the fact that operator ${\cal R}$ maps symmetries of
equation (\ref{qtt}) to symmetries. However we are using the notion of
formal recursion operator for a completely different aim.

Let ${\cal R}_{1}=A_{1}+B_{1}\,D_{t}$ and ${\cal R}_{2}=A_{2}+B_{2}\,D_{t}$
be two scalar formal recursion operators. Then the product
${\cal R}_{3}={\cal R}_{1} {\cal R}_{2}$, in which $D_{t}^{2}$ is replaced
by
$(M+ND_t)$ is also a scalar formal recursion operator whose components are
given by (\ref{***}).

An operator ${\cal S}=P+Q D_{t}$ is said to be implectic if
$$
{\cal F}^{*} S+\bar{\cal S} {\cal F}=0, \qquad \bar{\cal S}=\bar P+\bar Q D_{t}.
$$
Here and in the sequel the superscript $*$ denotes the adjoint
operator.  If $\cal S$ can be applied to symmetries, then it maps
symmetries to cosymmetries.  In the symmetry approach~$P$ and~$Q$ are
supposed to be formal non-commutative series with respect to $D_{x}$.

The operator equations of components of the formal implectic
operator~${\cal S}=P+Q D_{t}$
have the following form
\begin{equation} \label{eqcon1}
P_{tt}+U^{*} P_{t}+2 Q_{t} V+Q V_{t}=V^{*} P-P V-(QU+U^{*}Q)\,V-U_{t}^{*}P,
\end{equation}
\begin{equation} \label{eqcon2}
Q_{tt}+2 P_{t}+2 Q_{t} U+U^{*} Q_{t}=V^{*} Q-Q V-(QU+U^{*}Q)\,U-(PU+U^{*}P)-
(U_{t}^{*}Q+Q U_{t}).
\end{equation}
The linearization~$P+Q D_{t}$ of a variational derivative of any conserved
density for equation~(\ref{qtt}) satisfies equations (\ref{eqcon1}),
(\ref{eqcon2}) up to a ``small'' rest (see \cite{ss2}).

Let us consider equations of the form (\ref{qF}).
It follows from formulas (\ref{eqcon1}), (\ref{eqcon2}) that
equation (\ref{qF}) has no higher conservation laws. Moreover, it is
easy to prove that the density of any conservation law, up to total
derivatives, is of the form
$$
\rho=r_{1}(q,q_{x})\,q_{t}+r_{2}(q,q_{x}).
$$

All statements presented below can be easily reformulated for general
equations (\ref{qtt}).

{\bf Theorem 1.}
If equation (\ref{qF}) possesses an infinite sequence of
higher symmetries of the form
\begin{equation} \label{syms}
q_{\tau_{i}}=G_{i}(q,q_1,q_2,\dots , q_{r_{i}},\, q_{t},q_{t1}, q_{t2},
\dots,q_{ts_{i}})
\end{equation}
then there exists a formal recursion operator of the form
\begin{equation} \label{fro}
{\cal R}=(a_{0}+a_{-1} D^{-1}+\dots)+(D^{-1}+b_{-2}D^{-2}+\dots)\,D_{t},
\end{equation}
where $a_{i}, b_{i}$ are some functions of the variables (\ref{dynvar}).

For scalar evolution equations one can use (see \cite{ss, msy, mss}) the
residues
of powers of the formal recursion operator to derive the canonical
conservation laws
(\ref{cond}). Unfortunately for equations (\ref{qF}) this technique does not
work and we present a different way (cf.\cite{chi}) to get necessary
integrability
conditions (\ref{cond}).

Let $\cal R$ be a formal recursion operator of the form~(\ref{fro}). It is
then possible to find an operator
\begin{equation} \label{rm1}
{\cal R}^{-1}=(\alpha_{-1} D^{-1}+\alpha_{-2} D^{-2}\dots)+
(D^{-2}+\beta_{-3}D^{-3}+\dots)\,D_{t}
\end{equation}
such that ${\cal R}\,{\cal R}^{-1}={\cal R}^{-1}{\cal R}=1$. Recall that we
eliminate $D_{t}^{2}$ in virtue of ${\cal F}=0$ in the product of scalar
recursion operators. The operator ${\cal R}^{-1}$ is uniquely defined.

{\bf Theorem 2.} If~$\cal R$ is a formal recursion operator of the
form~(\ref{fro})  for equation~(\ref{qF}), then there is a unique
representation
of the total derivative operators $D_{x}$ and $D_{t}$ of the form
\begin{equation} \label{decomp}
D_{x}=\sum_{-\infty}^{2} \rho_{i} {\cal R}^{i}, \qquad
D_{t}=\sum_{-\infty}^{3} \sigma_{i} {\cal R}^{i}.
\end{equation}
Functions $\rho_{i}$ and $\sigma_{i}$ are densities and fluxes of some
(maybe trivial) conservation laws~(\ref{cond})
for equation (\ref{qF}).

The next formulas define five integrability conditions (\ref{cond})
for equations (\ref{qF}):
$$
\rho_{1}=u_{1},
$$
$$
\rho_{2}= v_{2}+\frac{2}{3} \sigma_{1},
$$
$$
\rho_{3}=6 \sigma_{2}-u_{1}\sigma_{1}+9 u_{0}-3 u_{1} v_{2}-\frac{1}{3}
u_{1}^{3},
$$
$$
\rho_{4}=6 \sigma_{3}-9 u_{1} \sigma_{2}+3 \sigma_{1}^{2}+27 u_{0}
u_{1}-u_{1}^{4}+81 v_{1}-
9 u_{1}^{2}v_{2}-27 v_{2}^{2}.
$$
$$
\rho_{5}=2 \sigma_{4}+18 \sigma_{1} \sigma_{2}-27 (\sigma_{1})_{t}-
3 \sigma_{1}^{2} u_{1}-3 \sigma_{3} u_{1}-9 \sigma_{1} u_{1} v_{2}-
\sigma_{1} u_{1}^{3}+27 \sigma _{1} u_{0}.
$$
The conditions mean that $\rho_{i}$ are densities of {\it local}
conservation laws for equation (\ref{qF}).  In other words, for any
$\rho_{i}$ there exists a corresponding function $\sigma_{i}$ depending
on variables~(\ref{dynvar}).

\subsection{List of integrable equations.}

{\bf Lemma 1}. If equation (\ref{qF}) satisfies conditions (\ref{cond}) with
$i=1,2,3,4$ then it is of the form
\begin{equation}
\begin{array}{l} \label{anzats}
q_{tt}=q_{xxx}+(A_{1}q_{t}+A_{2})\,q_{tx}+A_{3}\,q_{xx}^{2}+
(A_{4}q_{t}^{2}+A_{5} q_{t}+A_{6})\,q_{xx}+\\[3mm]
\qquad +A_{7} q_{t}^{4}+A_{8} q_{t}^{3}+A_{9} q_{t}^{2}+A_{10} q_{t}+A_{11},
\end{array}
\end{equation}
where the functions $A_{i}$ depend on $q$ and $q_{x}$ only.

It is easy to verify that the class of equations (\ref{anzats}) is invariant
with respect to point transformations $q \rightarrow \varphi(q)$. Moreover,
if all functions $A_{i}$ do not depend on $q$, then the shifts of the form
$q \rightarrow q+\lambda_{1} x+\lambda_{2} t,$ where $\lambda_{i}$ are
arbitrary constants, are also allowed.

{\bf Theorem 3.} Up to the transformations described above, any nonlinear
equation (\ref{qF}) satisfying integrability conditions with
$i=1,2,\dots,7$, coincides with the equation
\begin{equation} \label{new}
q_{tt}=q_{xxx}+(3 q_{x}+k)\, q_{xt}+(q_{t}-3 q_{x}^{2}-2 k q_{x}+6 {\cal
P})\,q_{xx}
-2 {\cal P}'\,q_{t}+6 {\cal P}'\,q_{x}^{2}+({\cal P}''+k {\cal P}')\,q_{x},
\end{equation}
where ${\cal P}(q)$ is any solution of equation of the form
\begin{equation} \label{wei}
 {\cal P}'^{2}=8 {\cal P}^{3}+k^{2} {\cal P}^{2}+
 c_{1} {\cal P}+c_{0}
\end{equation}
or with the equation
\begin{equation} \label{sh}
\displaystyle q_{tt}=q_{xxx}+ \left(\frac{3 q_{t}}{q_{x}}+
\frac{3}{2}\,X\right)\, q_{xt}-
\frac{1}{q_{x}} q_{xx}^{2}-
 \left(\frac{2 q_{t}^{2}}{q_{x}^{2}}+\frac{3 q_{t}}{2 q_{x}}\,X\right)\,
q_{xx}+
 c_{2}\Big(q_{x} q_{t}+ \frac{3}{2} q_{x}^{2}\,X\Big),
\end{equation}
where $X(q)=c_{2} q+c_{1}$ and $c_{i}$ are arbitrary constants.

{\bf Remarks}.  Actually, any equation (\ref{sh}) can be reduced to
the equation with $X(q)=const$ or to the equation with $X(q)=q$.  The
integrability conditions~6 and 7 have quite long expressions and we do not
present
them in the paper. We have only used these conditions to prove that
any
equation (\ref{qF}) for which all conservation laws (\ref{cond}) are
trivial is equivalent to a linear one.

\subsection{Recursion operators.}

In this Section we present a closed form of recursion operators for
models (\ref{new}) and (\ref{sh}). The existence of these recursion
operators implies the fact that all integrability conditions with $i\ge 1$
are fulfilled for these equations.

Let us consider equation (\ref{new}). This equation has the only non-trivial
conserved density given by
\begin{equation} \label{dens}
\rho=q_{t}-q_{x}^{2}+2 {\cal P}(q).
\end{equation}
The simplest higher symmetries of (\ref{new}) are
$$
q_{\tau}=q_{xx}+2 q_{x} q_{t}-q_{x}^{3}-k q_{x}^{2}+4 {\cal P} q_{x}
$$
and
$$
q_{\tau}=q_{xt}+q_{x} q_{xx}+q_{t}^{2}+(q_{x}^{2}+2 {\cal
P})\,q_{t}-q_{x}^{4}
-k q_{x}^{3}+6 {\cal P} q_{x}^{2}+({\cal P}'+k {\cal P})\,q_{x}
$$
They are generated by the following recursion operator
$$
{\cal R}=D+(q_{t}-2 q_{x}^{2}-k q_{x}+2 {\cal P})+q_{x} D^{-1}(D_{t}+
2 q_{xx}+2 {\cal P'})
$$
acting on the seed symmetries $q_{x}$ and $q_{t}$. A direct calculation
shows that this recursion operator satisfies (\ref{*}) and (\ref{**}). In
the degenerate case ${\cal P}\equiv 0,\, k=0$ we have
${\cal R}={\cal R}_{0}^{2},$ where ${\cal R}_0$ is defined by (\ref{r0}). The
operator $D_{t}+ 2 q_{xx}+2 {\cal P'}$ corresponds to the variational
derivative of the function (\ref{dens}) and therefore if we apply ${\cal R}$
to any local symmetry admitting the conservation law with density
(\ref{dens}), the result should be local.

Another recursion operator for (\ref{new}) has the form
$$
{\cal S}=D_{t}+(q_{xx}-q_{x}^{3}-k q_{x}^{2}+6 {\cal P} q_{x}+k {\cal P}+
{\cal P}')+q_{t} D^{-1}(D_{t}+2 q_{xx}+2 {\cal P'})
$$
One can verify that
$$
{\cal S}^{2}={\cal R}^{3}-k\,{\cal R}{\cal S}-\frac{c_1}{2}{\cal R}-c_{0}.
$$

For equation (\ref{sh}) a recursion operator is given by
$$
{\cal R}=\left(\frac{q_{t}}{q_{x}}+\frac{1}{2} X\right)-
q_{x} D^{-1}\left(\frac{1}{q_{x}} D_{t}+
\frac{q_{xt}}{q_{x}^{2}}-\frac{2 q_{t}
q_{xx}}{q_{x}^{3}}+\frac{c_{2}}{2}\right).
$$

The non-trivial conserved density for this equation is given by
\begin{equation} \label{dens1}
\rho=\frac{q_{t}}{q_{x}}+\frac{1}{2}X(q).
\end{equation}
In the case $X(q)=0, \, c_{1}=0$
the nonlocal variable $Q=D_{x}^{-1}(\rho)$ satisfies equation (\ref{311}).

\section{Linearization procedure.}

Each of equations (\ref{new}) and (\ref{sh}) has only one non-trivial
conserved local density.  The Burgers equation $u_{t}=u_{xx}+2 u
u_{x}$ possesses the same property: $D_{t}(u)=D_{x}(u_{x}+u^{2})$ is
the only conservation law for this equation.  The crucial step in the
linearization of the Burgers equation is to introduce the potential
$W$ of this conservation law.  By definition the variable $W$
satisfies conditions $W_{x}=u, \quad W_{t}=u_{x}+u^{2}$.  It is easy
to verify that the function $U=\exp{(W)}$ satisfies the heat equation
$U_{t}=U_{xx}$.

For equations (\ref{new}) and (\ref{sh}) the potentials of the conservation
laws also play a key role in linearization. However the procedure
of linearization is not so straightforward. To illustrate this procedure,
let us consider the simplest version $X=0$ of equation (\ref{sh}).
The potential of the conservation law for this equation
satisfies the conditions
$$
W_{x}=\frac{q_{t}}{q_{x}}, \qquad W_{t}=\frac{q_{xx}}{q_{x}}
+\frac{q_{t}^{2}}{q_{x}^{2}}.
$$
A simple computation shows that the equation admits the following non-local
symmetry $q_{y}=W$. Since
$$
D_{y}\left(\frac{q_{t}}{q_{x}}\right)=\frac{q_{xx}}{q_{x}^{2}},
$$
we have
$$
q_{yy}=W_{y}=-\frac{1}{q_{x}}, \qquad q_{yyy}=\frac{q_{t}}{q_{x}^{3}}
$$
or
$$
q_{x}=-\frac{1}{q_{yy}}, \qquad q_{t}=-\frac{q_{yyy}}{q_{yy}^{3}}.
$$
After the Legendre transformation
$$y=U_{z},\qquad q=U-z U_{z}$$
the latter pair of compatible equations becomes
$$U_{x}=U_{zz}, \qquad U_{t}=U_{zzz}.$$

For the more complicated equation
\begin{equation} \label{sh1}
\displaystyle q_{tt}=q_{xxx}+ \frac{3 q_{t}}{q_{x}}\,q_{xt}-
\frac{1}{q_{x}} q_{xx}^{2}-\frac{2 q_{t}^{2}}{q_{x}^{2}}\, q_{xx}+
c\, \Big(q_{xt}-\frac{q_{t}}{q_{x}} q_{xx}\Big)
\end{equation}
corresponding to the case $X=const\ne 0$ the same linearization scheme works
also.
The potential is defined by
$$
W_{x}=\frac{q_{t}}{q_{x}}, \qquad W_{t}=\frac{q_{xx}}{q_{x}}
+\frac{q_{t}^{2}}{q_{x}^{2}}+c\,\frac{q_{t}}{q_{x}},
$$
the non-local symmetry is given by $q_{y}=\exp{(-c W)}.$ It is not hard to
check that
$$
q_{x}=-c^{2}\,\frac{ q_{y}^{2}}{q_{yy}},
\qquad q_{t}=c^{3}\, \frac{q_{y}^{3}q_{yyy}-2 q_{y}^{2}
q_{yy}^{2}}{q_{yy}^{3}}.
$$
After a contact transformation
$$
y=\frac{1}{2} \exp{(-z)}\,(U_{z}+U), \qquad q=\frac{1}{2}
\exp{(z)}\,(U_{z}-U)
$$
these equations become
$$
U_{x}=\frac{c^{2}}{2}\,(U_{zz}-U), \qquad U_{t}=\frac{c^{3}}{4}\,(U_{zzz}
+U_{zz}-U_{z}-U).
$$

The most non-trivial case is $X=q$. In this case the potential $W$ is
defined by
$$
W_{x}=\frac{q_{t}}{q_{x}}+\frac{q}{2}, \qquad W_{t}=\frac{q_{xx}}{q_{x}}+
\frac{q_{t}^{2}}{q_{x}^{2}}+\frac{3 q_{t}}{2 q_{x}}\,q+\frac{3}{4} q^{2}
$$
but non-local symmetries of the form $q_{y}=F(q,W)$ do not exist. However
there exists a new non-local conservation law with potential $Z$ defined by
$$
Z_{x}=q^{2}-2 W q_{x}, \qquad Z_{t}=-\frac{1}{2} q^{3}-2 q_{x}-2 W q_{t}.
$$
Using these two potentials, we find a non-local symmetry
$ q_{y}=2 q \,\exp{(-\frac{Z}{4}-\frac{q W}{2})}.$ Expressing $q_{x}$ and
$q_{t}$ in terms of the $y$-derivative, we get
$$
q_{x}=-\frac{q^{3} q_{y}^{2}}{4 (q q_{yy}-2 q_{y}^{2})}, \qquad
q_{t}=\frac{q^{4} q_{y}^{3}(q^{2}q_{yyy}-9 q q_{y}q_{yy}+12 q_{y}^{3})}
{8 (q q_{yy}-2 q_{y}^{2})^{3}}.
$$
After the contact transformation $y=z+\frac{U}{U_{z}}$, $
q=-\frac{U_{z}}{U^{2}}$ we do not obtain linear equations but
$$
U_{x}=\frac{1}{4} D_{z}\left(\frac{U_{z}}{U^{2}}\right), \qquad
U_{t}=-\frac{1}{8} D_{z}\left(\frac{U_{zz}}{U^{3}}-\frac{3
U_{z}^{2}}{U^{4}}\right).
$$
To linearize the latter system one can introduce the potential $Y$ such that
$Y_{z}=U,\, Y_{x}=\frac{U_{z}}{4 U^{2}}$ and after that make a point
transformation $Y \leftrightarrow z$.

Equation (\ref{new}) can be linearized as follows. It is easy to verify that
it has a non-local symmetry $q_{y}=A(q) \exp{(-W)}$, where
$$
W_{x}=q_{t}-q_{x}^{2}+2 {\cal P}, \qquad W_{t}=q_{xx}-q_{x}^{3}+q_{t} q_{x}+
k (q_{t}-q_{x}^{2}+{\cal P})+6 {\cal P} q_{x}+{\cal P}'+w
$$
and
$$
B'^{2}=B^{4}+\frac{k^{2}}{2} B^{2}+8 w B+b_{0}, \qquad
B=-\frac{A'}{A}+\frac{k}{2}.
$$
It can be checked that
\begin{equation}\label{ee1}
q_{x}=\frac{q_{yy}}{q_{y}^{2}}+2 B(q),
\end{equation}
\begin{equation}\label{ee2}
q_{t}=-\frac{q_{yyy}}{q_{y}^{3}}+3 \frac{q_{yy}^{2}}{q_{y}^{4}}
+3 \frac{q_{yy}}{q_{y}^{2}} B(q)+
\frac{k}{2}\Big(\frac{q_{yy}}{q_{y}^{2}}+2 B(q)\Big)-
\frac{3}{2}\Big(B'(q)-B(q)^{2}\Big)
-\frac{k^{2}}{8}.
\end{equation}
Function ${\cal P}$ from equation (\ref{new}) is given by
$$
{\cal P}=-\frac{1}{4} B'+\frac{1}{4} B^{2}-\frac{k^{2}}{48},
$$
where the parameters of the elliptic functions ${\cal P}$ and $B$ are
related by
$$
c_{1}=\frac{k^{4}-16 b_{0}}{32},
\qquad c_{0}=w^{2}.
$$
After change of variables $z=q, u=y$ equations (\ref{ee1}), (\ref{ee2}) take
the following linear form
$$u_{x}=u_{zz}-2 B(z) u_{z},$$
$$
u_{t}=-u_{zzz}+3 B(z)\, u_{zz}+\frac{3}{2}\left(B'(z)-B(z)^{2}\right)\,
u_{z}+
\frac{k}{2}\Big(u_{zz}-2 B(z) u_{z}\Big)+\frac{k^{2}}{8}\,u_{z}.
$$

We see that in all cases there exists a non-local symmetry $q_{y}=G$
depending on the potentials such that $q_{x}$ and $q_{t}$ can be expressed
in terms of $y$-derivatives by formulas (\ref{linsys}). Using
this parametrization, one can construct particular solutions of equations
(\ref{new}) and~(\ref{sh}).

A parametrization of such kind arises not only for linearizable
equations but also for equations of KdV-type and associated linear spectral
problems. For example, let us consider the spectral problem
\begin{equation} \label{ls3}
\Psi_{xx}=(\lambda^3+u_1 \lambda^2+u_2 \lambda +u_3)\Psi
\end{equation}
For a non-local symmetry of this linear equation one can take (see
\cite{luis1,luis3})
 $$
   \bar \Psi_{yy}=\frac{\lambda}{a^2}\,\bar \Psi, \qquad \bar \Psi=
   \frac{1}{\sqrt{a}} \Psi.
 $$
 Then
 $$
   u_1=\frac{1}{4} a_y^2-\frac{1}{2}a a_{yy}, \qquad
   u_2=-a \Big(\log{(a)}\Big)_{xy}, \qquad
   u_3=-\frac{a_{xx}}{2 a}+\frac{3 a_x^2}{4 a^2},
 $$
 where $a(x,y)$ satisfies the Harry-Dym equation
 $$
   a_x=a^3 a_{yyy}.
 $$
At least on the local level the general solution of the latter equation
depends on three arbitrary functions $a_0(x)=a(x,0),\, a_1(x)=a_y(x.0),
\, a_2(x)=a_{yy}(x,0)$ of $x$ and therefore this parametrization
provides a generic potential in (\ref{ls3}).
Multi-phase solutions of the Harry-Dym equation lead to special
potentials of the spectral problem (\ref{ls3}).

\end{document}